\documentclass[aps,prd,twocolumn,showpacs,amsmath,amssymb]{revtex4-1}
\usepackage{amsmath}
\usepackage{graphicx}
\usepackage{subfigure}
\usepackage{epstopdf}
\usepackage{color}
\usepackage{multirow}
\usepackage{setspace}
\usepackage{overpic}
\usepackage{amssymb}
\usepackage[driverfallback=dvipdfmx, bookmarksnumbered, pdfstartview=FitH,colorlinks,urlcolor=blue, citecolor=blue,linkcolor=blue] {hyperref}
\usepackage{lineno}
\usepackage{bm}
\usepackage{rotating}
\usepackage[utf8]{inputenc}

\let\oldequation\equation
\let\oldendequation\endequation

\renewenvironment{equation}
  {\linenomathNonumbers\oldequation}
  {\oldendequation\endlinenomath}

\newcommand{\siga}{$45.8^{+9.0}_{-8.3}$}
\newcommand{\sigb}{$7.7^{+5.0}_{-4.3}$}

\newcommand{\nifia}{$7.0\sigma$}
\newcommand{\nifib}{$1.9\sigma$}

\newcommand{\bfa}{$[3.13^{+0.60}_{-0.56}({\rm stat}) \pm 0.09({\rm syst})] \times 10^{-4}$}

\newcommand{\bfastat}{[3.13^{+0.60}_{-0.56}({\rm stat})] \times 10^{-4}}
\newcommand{\bfbstat}{[1.84^{+1.19}_{-1.00}({\rm stat})] \times 10^{-4}}

\newcommand{\ratioa}{$(0.22\pm 0.04)\%$}

\newcommand{\tanca}  {$(0.75\pm 0.14)\tan^{4} \theta_C$}

\begin{document}

\title{\bf \boldmath
Measurement of the branching fraction of the doubly Cabibbo-suppressed decay $D^0\to K^+\pi^-\pi^0$ and search for $D^0\to K^+\pi^-\pi^0\pi^0$}

\author{
M.~Ablikim$^{1}$, M.~N.~Achasov$^{10,b}$, P.~Adlarson$^{68}$, M.~Albrecht$^{4}$, R.~Aliberti$^{28}$, A.~Amoroso$^{67A,67C}$, M.~R.~An$^{32}$, Q.~An$^{64,50}$, X.~H.~Bai$^{58}$, Y.~Bai$^{49}$
% ...
\\
\vspace{0.2cm}
(BESIII Collaboration)\\
\vspace{0.2cm} {\it
$^{1}$ Institute of High Energy Physics, Beijing 100049, People's Republic of China\\
% ...
$^{4}$ Bochum Ruhr-University, D-44780 Bochum, Germany\\
% ...
$^{10}$ G.I. Budker Institute of Nuclear Physics SB RAS (BINP), Novosibirsk 630090, Russia\\
% ...
$^{28}$ Johannes Gutenberg University of Mainz, Johann-Joachim-Becher-Weg 45, D-55099 Mainz, Germany\\
% ...
$^{32}$ Liaoning Normal University, Dalian 116029, People's Republic of China\\
% ...
$^{49}$ Southeast University, Nanjing 211100, People's Republic of China\\
$^{50}$ State Key Laboratory of Particle Detection and Electronics, Beijing 100049, Hefei 230026, People's Republic of China\\
% ...
$^{58}$ University of Jinan, Jinan 250022, People's Republic of China\\
% ...
$^{64}$ University of Science and Technology of China, Hefei 230026, People's Republic of China\\
% ...
$^{67}$ University of Turin and INFN, (A)University of Turin, I-10125, Turin, Italy; (B)University of Eastern Piedmont, I-15121, Alessandria, Italy; (C)INFN, I-10125, Turin, Italy\\
$^{68}$ Uppsala University, Box 516, SE-75120 Uppsala, Sweden\\
% ...
$^{b}$ Also at the Novosibirsk State University, Novosibirsk, 630090, Russia\\
% ...
}
}


\begin{abstract}
Using $2.93\,\rm fb^{-1}$ of $e^+e^-$ collision data collected at a center-of-mass energy of 3.773\,GeV with the BESIII detector, we present a measurement of the branching fraction of  the doubly Cabibbo-suppressed (DCS) decay $D^0\to K^+\pi^-\pi^0$ and a search for the DCS decay $D^0\to K^+\pi^-\pi^0\pi^0$.
The branching fraction of $D^0\to K^+\pi^-\pi^0$ is determined to be
$[3.13^{+0.60}_{-0.56}({\rm stat}) \pm 0.09({\rm syst})] \times 10^{-4}$.  No signal is observed for $D^0\to K^+\pi^-\pi^0\pi^0$ and an upper limit of $3.6 \times 10^{-4}$ is set on the branching fraction at the 90\% C.L.
We combine these results with the world-average branching fractions of their counterpart Cabibbo-favored decays
to determine the ratios of the doubly Cabibbo-suppressed over the Cabibbo-favored branching fractions, ${\mathcal B}(D^0\to K^+\pi^-\pi^0)/{\mathcal B}(D^0\to K^-\pi^+\pi^0)=(0.22\pm 0.04)\%$~and ${\mathcal B}(D^0\to K^+\pi^-\pi^0\pi^0)/{\mathcal B}(D^0\to K^-\pi^+\pi^0\pi^0)<0.40\%$ at the 90\% C.L., which
correspond to $(0.75\pm 0.14)\tan^{4} \theta_C$~and $1.37\times \tan^{4} \theta_C$, respectively, where $\theta_C$ is the Cabibbo angle.
\end{abstract}

\pacs{13.20.Fc, 14.40.Lb}

\maketitle

\oddsidemargin  -0.2cm
\evensidemargin -0.2cm

\section{Introduction}

Studies of doubly Cabibbo-suppressed (DCS) decays of charmed mesons provide important information on charmed-hadron dynamics. 
The ratio of the branching fraction of a given DCS $D^{0(+)}$ decay relative to its Cabibbo-favored (CF) counterpart is naively expected to be about $(0.5-2)\times {\rm tan}^4\theta_C$ ($\theta_C$ is the Cabibbo mixing angle)~\cite{Lipkin,theory_1}.
Recently, BESIII reported the observation of the DCS decay $D^+\to K^+\pi^+\pi^-\pi^0$~\cite{bes3_DCS_Kpipipi0,bes3-DCS-Dp-K3pi-v2} (charge conjugate processes are implied throughout this paper).
The branching fraction of this decay averaged over the two measurements reported in Refs.~\cite{bes3_DCS_Kpipipi0,bes3-DCS-Dp-K3pi-v2} is $[1.13 \pm 0.08({\rm stat}) \pm 0.03({\rm syst})]\times 10^{-3}$,
which gives a DCS/CF branching fraction ratio of $(6.28\pm0.52)\tan^4\theta_C$.
Comprehensive measurements of the DCS decays of other charmed mesons, especially for isospin symmetrical decays of $D^0$,
may shed light on the origin of this anomalously large DCS/CF branching fraction ratio.

So far, only a few DCS $D^0$ decays, namely $D^0\to K^+\pi^-$, $D^0\to K^+\pi^-\pi^0$ and $D^0\to K^+\pi^-\pi^-\pi^+$, have been observed, with decay branching fractions extracted from the ratio of DCS/CF decay branching fractions from the experiments determining $D^0$-$\bar D^0$ mixing parameters or coherence parameters~\cite{pdg2020}.
In this paper, we present the first direct measurements of the branching fractions of $D^0\to K^+\pi^-\pi^0$ and $D^0\to K^+\pi^-\pi^0\pi^0$
by analyzing 2.93\,fb$^{-1}$ of $e^+e^-$ collision data~\cite{lum_bes3} taken at a center-of-mass energy of 3.773~GeV with the BESIII detector.
Because the traditional hadronic tag method suffers from complex quantum-correlation effects~\cite{zzxing},
this analysis is performed with the semileptonic tag method adopted in our previous work~\cite{bes3-DCS-Dp-K3pi-v2}.
Our direct measurements would benefit the constraint of the charm mixing parameters when combining with individual CF $D^0$ decay branching fraction.

\section{Data and Monte Carlo}

The BESIII detector is a magnetic
spectrometer~\cite{BESIII} located at the Beijing Electron
Positron Collider (BEPCII)~\cite{Yu:IPAC2016-TUYA01}. The
cylindrical core of the BESIII detector consists of a helium-based
 multilayer drift chamber (${\rm MDC}$), a plastic scintillator time-of-flight
system (${\rm TOF}$), and a CsI(Tl) electromagnetic calorimeter (${\rm EMC}$),
which are all enclosed in a superconducting solenoidal magnet
providing a 1.0~T magnetic field. The solenoid is supported by an
octagonal flux-return yoke with resistive plate counter muon-identifier modules interleaved with steel. The acceptance of
charged particles and photons is 93\% over the $4\pi$ solid angle. The
charged-particle momentum resolution at $1~{\rm GeV}/c$ is
$0.5\%$, and the resolution of specific ionization energy loss~(d$E$/d$x$) is $6\%$ for electrons
from Bhabha scattering. The EMC measures photon energies with a
resolution of $2.5\%$ ($5\%$) at $1$~GeV in the barrel (end-cap)
region. The time resolution of the TOF barrel part is 68~ps, while
that of the end-cap part is 110~ps.
Details about the design and performance of the BESIII detector are given in Ref.~\cite{BESIII}.

Simulated samples produced with the {\sc geant4}-based~\cite{geant4} Monte Carlo (MC) package, which
includes the geometric description of the BESIII detector and the
detector response, are used to determine the detection efficiency
and to estimate the backgrounds. The simulation includes the beam-energy spread and initial-state radiation in the $e^+e^-$
annihilations modeled with the generator {\sc kkmc}~\cite{kkmc}.
The inclusive MC samples consist of the production of $D\bar{D}$ pairs,
the non-$D\bar{D}$ decays of the $\psi(3770)$, the initial-state radiation
production of the $J/\psi$ and $\psi(3686)$ states, and the
continuum processes.
The known decay modes are modelled with {\sc
evtgen}~\cite{evtgen} using the branching fractions taken from the
Particle Data Group (PDG)~\cite{pdg2020}, and the remaining unknown decays of the charmonium states are
modeled by {\sc lundcharm}~\cite{lundcharm}. Final-state radiation is incorporated using the {\sc photos} package~\cite{photos}.

The $D^0\to K^+\pi^-\pi^0$ decay is simulated using an MC generator which combines the resonant decays $D^0\to K^*(892)^0\pi^0$, $D^0\to K^*(892)^+\pi^-$, $D^0\to K^+\rho(770)^-$, and a three-body phase-space model.
The $D^0\to K^+\pi^-\pi^0\pi^0$ decay is simulated with a four-body phase-space model.
The $D^0\to K^-e^+\nu_e$ decay is simulated with the modified pole model~\cite{MPM}
with the pole mass fixed at the $D_s^{*+}$ nominal mass~\cite{pdg2020} and the other parameters quoted from \cite{bes3-D0-kev}.

\section{Measurement method}

The center-of-mass energy of 3.773 GeV lies above the $D\bar D$ production threshold but below that of $D^*\bar D$.
At this energy point, the $D^0\bar D^0$ pairs are produced copiously and are not accompanied by additional hadrons.
This allows $D$ decays to be studied with the double-tag method.
In this analysis double-tag events refer to those in which the DCS decays $D^0\to K^+\pi^-\pi^0$ or $D^0\to K^+\pi^-\pi^0\pi^0$ are found on the recoiling side of the semileptonic decay $\bar D^0\to K^+ e^-\bar \nu_e$.
The branching fraction of $D^0\to K^+\pi^-\pi^0$ or $D^0\to K^+\pi^-\pi^0\pi^0$ is determined by
\begin{equation}\label{equ:br}
{\mathcal B}_{{\rm DCS}} = \frac{N_{\rm DT}} {2\cdot N_{D^0\bar D^0}\cdot
\epsilon_{\rm DT}\cdot  {\mathcal B}_{\rm SL}},
\end{equation}
where
$N_{D^0\bar D^0}=(10597\pm28\pm98)\times 10^3$ is the total number of $D^0\bar D^0$ pairs in the data sample determined in our previous work~\cite{bes3-crsDD},
$N_{\rm DT}$ is the signal yield of the double-tag events obtained from the data sample,
$\epsilon_{\rm DT}$ is the effective efficiency of reconstructing the double-tag events,
and
${\mathcal B}_{\rm SL}$ is the branching fraction of the semileptonic decay $\bar D^0\to K^+ e^-\bar \nu_e$ taken from the PDG~\cite{pdg2020}.

\section{Event selection}

The double-tag candidates are required to contain at least two good photons for $D^0\to K^+\pi^-\pi^0$ and four for $D^0\to K^+\pi^-\pi^0\pi^0$ as well as exactly four charged tracks for both modes.
We use the same selection criteria for $K^\pm$, $\pi^\pm$, $e^-$, and $\pi^0$ candidates as were used in our previous studies~\cite{bes3_DCS_Kpipipi0,epjc76,cpc40,bes3-Dp-K1ev,bes3-D-b1enu}.
All charged tracks are required to originate from a region within $|\rm{cos\theta}|<0.93$, $|V_{xy}|<$ 1\,cm and $|V_{z}|<$ 10\,cm.
Here, $\theta$ is the polar angle of the charged track with respect to the MDC axis, $|V_{xy}|$ and $|V_{z}|$ are the distances of closest approach of the charged track to the interaction point perpendicular to and along the MDC axis, respectively.
Particle identification (PID) of kaons and pions is performed with the combined d$E$/d$x$ and TOF information to calculate their corresponding confidence levels.
Charged tracks with confidence level for kaon (pion) hypothesis greater than that for pion (kaon) hypothesis are assigned as kaon (pion) candidates.

Photon candidates are selected by using the information recorded by the EMC. The shower time is required to be within 700\,ns of the event start time. The shower energy is required to be greater than 25 (50)\,MeV if the crystal with the maximum deposited energy in that cluster is in the barrel~(end-cap) region~\cite{BESIII}. The opening angle between the shower direction and the extrapolated position on the EMC of the closest  charged track must be greater than $10^{\circ}$.
The $\pi^0$ candidates are formed by photon pairs with invariant mass within $(0.115,\,0.150)$\,GeV$/c^{2}$. To improve the resolution, a kinematic fit constraining the $\gamma\gamma$
invariant mass to the $\pi^{0}$ known mass~\cite{pdg2020} is imposed on the selected photon pair.

In the selection of the $D^0\to K^+\pi^-\pi^0\pi^0$ candidates,
the invariant mass of the $\pi^0\pi^0$ combination is required to be outside of the interval $(0.388,0.588)$\,GeV/$c^2$ to reject the dominant peaking background from the singly Cabibbo-suppressed decay $D^0\to K^+\pi^-K_S^0(\to\pi^0\pi^0)$. This requirement corresponds to about five standard deviations of the experimental $K_S^0$ mass resolution.
The signal candidates for $D^0\to K^+\pi^-\pi^0$ or $D^0\to K^+\pi^-\pi^0\pi^0$ are identified with two variables: the energy difference
\begin{equation}
\Delta E \equiv E_{D^0} - E_{\rm beam}
\label{eq:deltaE}
\end{equation}
and the beam-constrained mass
\begin{equation}
M_{\rm BC} \equiv \sqrt{E^{2}_{\rm beam}-|\vec{p}_{D^0}|^{2}}.
\label{eq:mBC}
\end{equation}
Here, $E_{\rm beam}$ is the beam energy, $\vec{p}_{D^0}$ and $E_{D^0}$ are the momentum and energy of the $D^0$ candidate in the $e^+e^-$ rest frame, respectively.
If there are multiple candidates for the hadronic side,
only the one with the minimum $|\Delta E|$ is kept.
The correctly reconstructed $D^0$ candidates concentrate around zero in the $\Delta E$ distribution
and around the  $D^0$ nominal mass in the $M_{\rm BC}$ distribution.
The events satisfying $\Delta E\in(-54,40)$\,MeV for $D^0\to K^+\pi^-\pi^0$ and $\Delta E\in(-60,30)$\,MeV for $D^0\to K^+\pi^-\pi^0\pi^0$ are kept for further analysis.

After the hadronic $D^0$ mesons are reconstructed, the candidates for $\bar D^0\to K^+e^-\bar \nu_{e}$ are selected from the remaining tracks that have not been used to select the hadronic side.
Then, the number of extra charged tracks ($N_{\rm extra}^{\rm charge}$) is required to be zero.
The charge of the electron candidate is required to be opposite to that of the kaon from the hadronic $D^0$ decay.
Electron PID uses the combined d$E$/d$x$, TOF, and EMC information, with which the combined confidence levels under the electron, pion, and kaon hypotheses ($CL_e$, $CL_{\pi}$, and $CL_{K}$) are calculated.
Electron candidates are required to satisfy $CL_e>0.001$ and $CL_e/(CL_e+CL_\pi+CL_K)>0.8$.
To reduce the background due to mis-identification between hadrons and electrons, the energy of the electron candidate deposited in the EMC is further required to be greater than 0.8 times its measured momentum.
Then, to partially compensate the effects of final-state radiation and \textsl{bremsstrahlung} (FSR recovery), the four-momenta of photon(s) within $5^\circ$ of the initial electron direction are added to the electron four-momentum measured by the MDC.

The charged kaons from the semileptonic decay are required to satisfy the same PID criteria as the kaons from the hadronic decays, and to have a charge opposite to that of the electron.
To suppress potential backgrounds from hadronic decays with a misidentified electron,
the invariant mass of the $K^+e^-$ combination, $M_{K^+e^-}$, is required to be less than 1.8~GeV/$c^2$.
Furthermore, we require that
the maximum energy of extra photons ($E^{\rm max}_{\rm extra\,\gamma}$) which have not been used in the tag
selection is less than 0.25~GeV and there is no extra $\pi^0$ candidate ($N_{\rm extra\,\pi^0}$).

The semileptonic $\bar D^0$ decay is identified using a kinematic quantity defined as
\begin{equation}
U_{\mathrm{miss}}\equiv E_{\mathrm{miss}}-|\vec{p}_{\mathrm{miss}}|.
\end{equation}
Here, $E_{\mathrm{miss}}\equiv E_{\mathrm{beam}}-E_{K^+}-E_{e^-}$ and $\vec{p}_{\mathrm{miss}}\equiv
\vec{p}_{\bar D^0}-\vec{p}_{K^+}-\vec{p}_{e^-}$ are the missing energy and momentum of the double-tag event in the $e^+e^-$ center-of-mass system, in
which $E_{K^+}$ and $\vec{p}_{K^+}$ are the energy and momentum of the $K^+$, and
$E_{e^-}$ and $\vec{p}_{e^-}$ are the energy and momentum of the $e^-$ candidate. The
$U_{\mathrm{miss}}$ resolution is improved by constraining the $D^0$ energy to the beam energy and $\vec{p}_{\bar D^0} \equiv {-\hat{p}_{D^0}}\cdot\sqrt{E_{\mathrm{beam}}^{2}-M_{D^0}^{2}}$, where $\hat{p}_{D^0}$ is the unit vector in the momentum direction of the $D^0$ and $M_{D^0}$ is the $D^0$ nominal mass~\cite{pdg2020}.

\begin{figure}[htbp]
\centering
\includegraphics[width=0.5\textwidth]{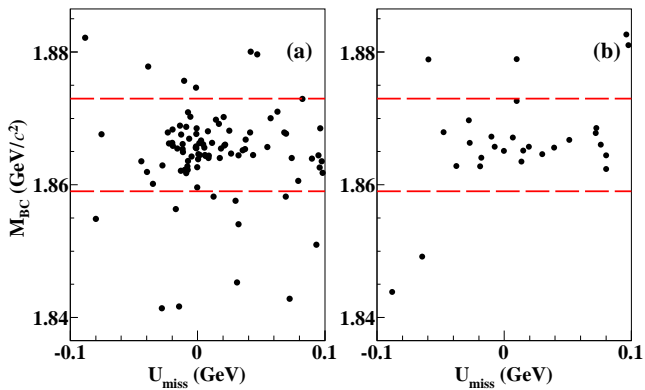}
\caption{Distributions of $M_{\rm BC}$ versus $U_{\mathrm{miss}}$ of the accepted double-tag candidate events for (a) $D^0\to K^+\pi^-\pi^0$ and (b) $D^0\to K^+\pi^-\pi^0\pi^0$ versus $\bar D^0\to K^+ e^-\bar \nu_e$ decays in data. The area between dashed red lines show the $M_{\rm BC}$ signal region.}
\label{fig:M_BC_M2}
\end{figure}

\begin{figure}[htbp]
  \centering
  \includegraphics[width=0.5\textwidth]{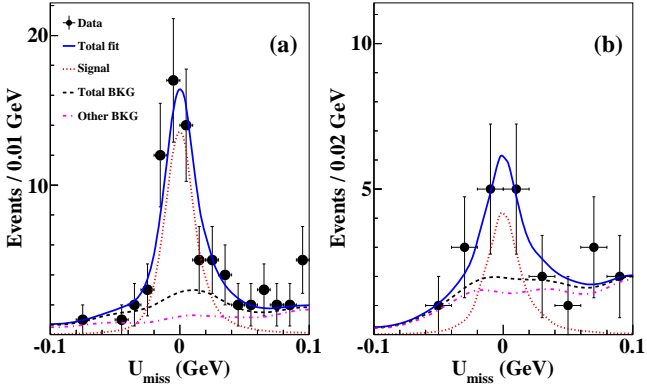}
\caption{Fits to the $U_{\mathrm{miss}}$ distributions of the accepted double-tag candidate events for (a) $D^0\to K^+\pi^-\pi^0$ and (b) $D^0\to K^+\pi^-\pi^0\pi^0$ versus $\bar D^0\to K^+ e^-\bar \nu_e$ decays.
The points with error bars are data.
The blue solid curves are the total fit results (Total fit). The red dotted and black dashed curves are the fitted signal (Signal) and background (Total BKG) components, respectively.
The component between the black dashed and pink dot-dashed curves is the peaking background and the pink dot-dashed curve represents the other background contributions (Other BKG).
}
\label{fig:fits_umiss}
\end{figure}

Figure \ref{fig:M_BC_M2} shows the distributions of $M_{\rm BC}$ versus $U_{\rm miss}$ of the double-tag candidate events in data. The clusters around the known $D^0$ mass along the $y$ axis and zero along the $x$ axis are the signal double-tag candidate events.
The signal region is selected around the known $D^0$ mass: those candidates satisfying $M_{\rm BC}\in (1.859,1.873)$\,GeV/$c^2$ are kept for further analysis.
After the implementation of the above-mentioned requirements, the $U_{\rm miss}$ distributions of the surviving events are shown in Fig.~\ref{fig:fits_umiss}.

The detection efficiencies $\epsilon_{{\rm DT}}$ obtained from signal MC samples are $(19.49\pm0.14)\%$ and $(5.56\pm0.07)\%$ for the double-tag events
of $D^0\to K^+\pi^-\pi^0$ and $D^0\to K^+\pi^-\pi^0\pi^0$ versus $\bar D^0\to K^+e^-\bar \nu_e$, respectively,
where the efficiencies include the branching fraction of $\pi^0\to \gamma\gamma$ and the uncertainties are statistical only.

The background components and corresponding ratios in the total background are described below.
For $D^0\to K^+\pi^-\pi^0$ versus $\bar  D^0\to K^+ e^- \bar{\nu}_{e}$,
the peaking backgrounds are mainly from the CF modes
$D^0\to K^- \pi^+\pi^0$ versus $\bar  D^0\to K^+ e^- \bar{\nu}_{e}$ due to
the mis-identification between kaons and pions in the hadronic side (36.0\%) and
the mis-identification between kaons and electrons in the hadronic side  (12.9\%);
while the residual backgrounds are
$D^0\to K^- \pi^+\pi^0$ versus $\bar D^0\to K^+\pi^-\pi^0$ (7.9\%),
$D^0\to K^+ K^-$ versus $\bar D ^0\to K^+ e^- \bar{\nu}_{e}$ (6.5\%),
$D^0\to \bar K^0 \pi^+\pi^-$ versus $\bar D^0\to K^+ e^- \bar{\nu}_{e}$ (5.8\%) and
other decay modes (30.9\%).
For $D^0\to K^+\pi^-\pi^0\pi^0$ versus $\bar  D^0\to K^+ e^- \bar{\nu}_{e}$,
the peaking backgrounds are mainly from
$D^0\to K^- e^+ \nu_{e}$ versus $\bar{D}^0\to K^+\pi^-\pi^0\pi^0$ (30.6\%);
while the residual backgrounds are
$D^{0}\to K_{S}^{0} \pi^+ \pi^- \pi^0$ versus $\bar{D}^0\to K^+ e^- \bar{\nu}_{e}$(8.2\%),
$D^{0}\to K^+ K^- \pi^0$ versus $\bar{D}^0\to K^+ e^- \bar{\nu}_{e}$ (8.2\%),
$D^{0}\to K_{L}^{0} \pi^+\pi^-$ versus $\bar{D}^0\to K^+ e^- \bar{\nu}_{e}$ (4.1\%),
and other decay modes (49.0\%).

To measure the signal yields, unbinned maximum-likelihood fits are performed on the $U_{\mathrm{miss}}$ distributions.
The non-peaking backgrounds (including a small contribution from wrongly reconstructed semileptonic candidates) are described by the corresponding MC-simulated shapes.
The background shapes are derived from the inclusive MC sample and the
signal shapes from the signal MC samples.
The yield of the peaking background is fixed based on the known branching fractions and the mis-identification rates, and the yields of the signal and
non-peaking backgrounds are free parameters of the fits.
The fit results are shown in Fig.~\ref{fig:fits_umiss}.
From these fits, we measure \siga~signal events for the decay $D^0\to K^+\pi^-\pi^0$ and \sigb~signal events for $D^0\to K^+\pi^-\pi^0\pi^0$.
These results give the product branching fractions to be
${\mathcal B}(D^0\to K^+\pi^-\pi^0)\cdot{\mathcal B}(\bar D^0\to K^+ e^-\bar \nu_\nu)=[1.11^{+0.21}_{-0.20}(\rm stat)] \times 10^{-5}$,
and
${\mathcal B}(D^0\to K^+\pi^-\pi^0\pi^0)\cdot{\mathcal B}(\bar D^0\to K^+ e^-\bar \nu_\nu)=[6.53^{+4.24}_{-3.65}(\rm stat)] \times 10^{-6}$.
Combining the world average of ${\mathcal B}(\bar D^0\to K^+ e^-\bar \nu_\nu)=(3.541\pm0.034)\%$~\cite{pdg2020},
we obtain
\begin{equation}
{\mathcal B}(D^0\to K^+\pi^-\pi^0)=\bfastat,
\nonumber
\end{equation}
and
\begin{equation}
{\mathcal B}(D^0\to K^+\pi^-\pi^0\pi^0)=\bfbstat.
\nonumber
\end{equation}
The statistical
significance of the signal is calculated by $\sqrt{-2{\rm ln ({\mathcal L_0}/{\mathcal
      L_{\rm max}}})}$, where
${\mathcal L}_{\rm max}$ and ${\mathcal L}_0$ are the maximal
likelihood of the fits with and without the signal contribution, respectively.
These significances are determined to be \nifia~and \nifib~for $D^0\to K^+\pi^-\pi^0$ and $D^0\to K^+\pi^-\pi^0\pi^0$, respectively.

\begin{figure}[htbp]
\centering
\includegraphics[width=0.5\textwidth]{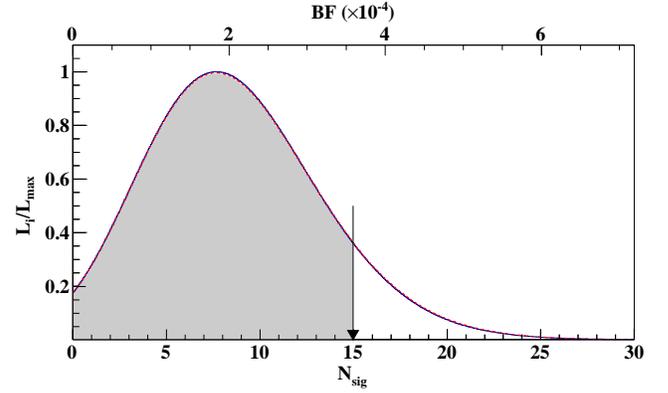}
\caption{Distributions of normalized likelihood distributions versus the signal yield $N_{\rm sig}$ and branching fraction of $D^0\to K^+\pi^-\pi^0\pi^0$.
The results obtained with and without incorporating the systematic uncertainty are shown by the red dashed and blue solid curves, respectively. The black arrow shows the result corresponding to 90\% confidence level.}
\label{fig:upper}
\end{figure}

The upper limit on the branching fraction of the decay $D^0\to K^+ \pi^- \pi^0 \pi^0$ is determined to be $3.6\times 10^{-4}$ at 90\% confidence level, using the Bayesian approach~\cite{UPM} after incorporating the systematic uncertainty.
The distribution of the likelihood versus the assumed branching fraction is shown in Fig.~\ref{fig:upper}.

\section{Systematic uncertainties}


The systematic uncertainties originating from $e^-$ tracking (PID) efficiencies
are studied by using a control sample of $e^+e^-\to\gamma e^+ e^-$ events.
The efficiency ratios of data and MC simulation for $e^-$ tracking and $e^-$ PID are
$(101.0\pm0.2)\%$ and $(101.2\pm0.2)\%$, respectively.
Here, the two dimensional (momentum and $\cos\theta$) $e^-$ tracking (PID) efficiencies from the control sample have been re-weighted to match those in the signal decays.
The systematic uncertainties associated with the $K^+$ and $\pi^-$ tracking (PID) efficiencies are investigated with $D^0 \to K^-\pi^+$, $K^-\pi^+\pi^0$, $K^-\pi^+\pi^+\pi^-$ versus~$\bar D^0 \to K^+\pi^-$, $K^+\pi^-\pi^0$, $K^+\pi^-\pi^-\pi^+$, as well as $D^+\to K^- \pi^+\pi^+$ versus~$D^-\to K^+\pi^-\pi^-$ double-tag hadronic $D\bar D$ events, using a sample with a missing $K^{+}$ or $\pi^{-}$.
The ratios of tracking or PID efficiencies for charged kaons and pions between data and MC simulation are listed in Table~\ref{correct}.
Here, the momentum dependent $K^+(\pi^-)$ tracking (PID) efficiencies from control samples have been re-weighted to match those in the signal decays.
After correcting the signal MC efficiencies by these factors, the residual uncertainties on the tracking (PID) efficiencies of
$e^-$, $K^+$, and $\pi^-$ are assigned as 0.2\%\,(0.2\%), 0.3\%\,(0.2\%), and 0.2\%\,(0.2\%), respectively.

\begin{table}[htp]
\centering
\caption{The ratios of efficiencies of $K^+$ tracking, $K^+$ PID, $\pi^-$ tracking, and $\pi^-$ PID between data and MC simulation.}
\label{correct}
\centering
\begin{tabular}{ccc}
  \hline \hline
Source &$D^0\to K^+\pi^-\pi^0$ (\%) & $D^0\to K^+\pi^-\pi^0\pi^0$ (\%) \\ \hline
$K^+_{\rm tracking}$         &$101.1 \pm 0.3$   &$101.7 \pm 0.3$  \\
$K^+_{\rm PID}$              &$100.0 \pm 0.2$   &$100.0 \pm 0.2$  \\
$\pi^-_{\rm tracking}$       &$100.1 \pm 0.2$   &$100.2 \pm 0.2$  \\
$\pi^-_{\rm PID}$             &$99.6 \pm 0.2$    &$99.8 \pm 0.2$ \\
\hline
\end{tabular}
\end{table}

The systematic uncertainty of $\pi^0$ reconstruction efficiency is investigated by using the double-tag hadronic $D\bar D$  decays of
$\bar D^0\to K^+\pi^-\pi^0$ and $\bar D^0\to K^0_S\pi^0$ tagged by either $D^0\to K^-\pi^+$ or $D^0\to K^-\pi^+\pi^+\pi^-$~\cite{epjc76,cpc40}.
The systematic uncertainty on the $\pi^0$ reconstruction efficiency is assigned as 0.8\% for each $\pi^0$.

The systematic uncertainty associated with the $U_{\rm miss}$ fit is estimated by comparing the baseline branching-fraction result with the result obtained
with alternative signal shapes and background shapes.
The systematic uncertainty due to the assumed signal shape is estimated by replacing the nominal description with one convolved with a Gaussian resolution function.
Here, the parameters used in the convolved Gaussian function representing the data-MC simulation difference are
obtained from the CF decay $D^0\to K^-\pi^+\pi^0(\pi^0)$.
The change in the branching fraction due to the assumed signal shape is found to be negligible.
The systematic uncertainty from the simulated background shape is taken into account by varying the
dominant peaking background component by $\pm 1\sigma$.
The change in the re-measured branching fraction, 1.0\%, is assigned as the systematic uncertainty associated with the background shape for $D^0\to K^+\pi^-\pi^0$ decays, while that for $D^0\to K^+\pi^-\pi^0\pi^0$ decays, is found to be negligible.
In addition, the effects of other background sources, examined by varying their size and shape, are also negligible.

The systematic uncertainties due to the requirements of $\Delta E$ and $M_{\rm BC}$ for the hadronic side as well
as the requirement of $M_{K^+e^-}$ for the semileptonic side are studied
by using control samples of the CF decay $D^0\to K^-\pi^+\pi^0(\pi^0)$ versus $\bar D^0\to K^+e^-\bar \nu_e$.
The corresponding uncertainties are taken to be the differences of the acceptance efficiencies between data and MC simulation. These uncertainties are all found to be 0.1\%.
The systematic uncertainty associated with the $K^0_S$ veto in the $M_{\pi^0\pi^0}$ distribution is assigned by varying the mass window by $\pm20$~MeV/$c^2$.
The maximum relative change in the measured branching fraction is not significantly larger than the statistical
uncertainty after considering the correlations between the signal yields, hence this uncertainty is ignored~\cite{ksocut}.


The systematic uncertainty due to MC modeling is assigned to be the difference between
the nominal efficiency and the average efficiency based on the signal MC events of the various components.
Besides individual phase-space decays, the resonant decays $D^0\to K^*(892)^0\pi^0$, $D^0\to K^*(892)^+\pi^-$, and $D^0\to K^+\rho(770)^-$
have been considered for $D^0\to K^+\pi^-\pi^0$;
and the resonant decays $D^0\to K^*(892)^0\pi^0\pi^0$, $D^0\to K^*(892)^+\pi^-\pi^0$, and $D^0\to K^+\pi^0\rho^-$ have been considered for $D^0\to K^+\pi^-\pi^0\pi^0$.
The corresponding systematic uncertainties are assigned as 1.9\% and 3.6\% for $D^0\to K^+\pi^-\pi^0$ and $D^0\to K^+\pi^-\pi^0\pi^0$, respectively.
The uncertainty in the MC modeling of the semileptonic decay of $\bar D^0\to K^+e^-\bar \nu_e$ has been estimated in our previous work and is negligible~\cite{bes3-D0-kev}.

The systematic uncertainty due to the $E^{\rm max}_{\rm extra\, \gamma}$, $N_{\rm extra}^{\rm charge}$, and $N_{\rm extra\,\pi^0}$ requirements
is estimated by using a control sample of the CF decay $D^0\to K^-\pi^+\pi^0(\pi^0)$ versus $\bar D^0\to K^+ e^-\bar \nu_e$.
The differences in the acceptance efficiencies between data and MC simulation, 0.2\% and 0.8\%, are taken as the corresponding systematic uncertainties for the $D^0\to K^+\pi^-\pi^0$ and $D^0\to K^+\pi^-\pi^0\pi^0$ decays, respectively.

The uncertainties due to MC sample sizes are 0.7\% and 1.2\% for $D^0\to K^+\pi^-\pi^0$ and $D^0\to K^+\pi^-\pi^0\pi^0$ decays, respectively.
The uncertainty from FSR recovery is estimated as 0.3\% as in $\bar D^0\to K^+ e^-\bar \nu_e$ decays~\cite{bes3-D0-kev}.
The total number of the $D^0\bar D^0$ pairs in the data sample is cited from Ref.~\cite{bes3-crsDD} and is known with a precision that induces a systematic uncertainty of 0.9\%.
The branching fraction of $\bar D^0\to K^+e^-\bar \nu_e$ contributes a systematic uncertainty of  1.0\%~\cite{pdg2020}.

Adding all these uncertainties in quadrature yields a total systematic uncertainty of
2.9\% for $D^0\to K^+\pi^-\pi^0$ and 4.0\% for $D^0\to K^+\pi^-\pi^0\pi^0$.
The systematic uncertainties discussed above are summarized in Table~\ref{tab:relsysuncertainties}.

\begin{table}[htpb]
\centering
\caption{Systematic uncertainties (in \%) in the determination of the branching fractions.}
\label{tab:relsysuncertainties}
\centering
\begin{tabular}{ccc}
  \hline
  \hline
Source                                     & $K^+\pi^-\pi^0$ & $K^+\pi^-\pi^0\pi^0$ \\ \hline
Tracking of $K^+$, $e^-$, and $\pi^-$& 0.7 & 0.7 \\
PID of $K^+$, $e^-$, and $\pi^-$     & 0.5 & 0.5 \\
$\pi^0$ reconstruction                     & 0.8 & 1.6 \\
$K_{S}^{0}$ veto                           & N/A & Ignored \\
MC model                                   & 1.9 & 3.6 \\
$U_{\rm miss}$ fit                         & 1.0 & Negligible \\
$\Delta E$ requirement                     & 0.1 & 0.1 \\
$M_{\rm BC}$ requirement                     & 0.1 & 0.1 \\
$E^{\rm max}_{\rm extra\,\gamma}\&N_{\rm extra\,\pi^0}\&N_{\rm extra}^{\rm charge}$  & 0.2 & 0.8\\
MC statistics                              & 0.7 & 1.2 \\
FSR recovery                               & 0.3 & 0.3 \\
$N_{D^0\bar D^0}$                          & 0.9 & 0.9 \\
Quoted branching fraction                 & 1.0 & 1.0 \\ \hline
Total                                      & 2.9 & 4.0 \\
\hline
\end{tabular}
\end{table}

\section{Summary}

In conclusion, using $2.93\,\rm fb^{-1}$ of $e^+e^-$ collision data accumulated at a center-of-mass energy of 3.773 GeV with the BESIII detector,
we have measured the branching fraction of the DCS decay of $D^0\to K^+\pi^-\pi^0$ and performed a search for the DCS decay $D^0\to K^+\pi^-\pi^0\pi^0$.
The branching fraction of  $D^0\to K^+\pi^-\pi^0$ is determined to be \bfa, which
is consistent with the PDG value~\cite{pdg2020}.
No significant signal is seen for  $D^0\to K^+\pi^-\pi^0\pi^0$ and an upper limit of $3.6 \times 10^{-4}$ is set on the branching fraction at the 90\% C.L.
Using the world-average value of ${\mathcal B}(D^0\to K^-\pi^+\pi^0)=(14.4\pm0.5)\%$~\cite{pdg2020},
we obtain the DCS/CF ratio ${\mathcal B}(D^0\to K^+\pi^-\pi^0)/{\mathcal B}(D^0\to K^-\pi^+\pi^0)=$ \ratioa, corresponding to \tanca.
Our result for  $D^0\to K^+\pi^-\pi^0\pi^0$ and the world-average value of  ${\mathcal B}(D^0\to K^-\pi^+\pi^0\pi^0)=(8.86\pm0.23)\%$~\cite{pdg2020} leads to the upper limit ${\mathcal B}(D^0\to K^+\pi^-\pi^0\pi^0)/{\mathcal B}(D^0\to K^-\pi^+\pi^0\pi^0) < 0.40\%$ at the 90\% C.L., corresponding to $1.37\times \tan^{4} \theta_C$.
In the future, amplitude analyses of these two decays with larger data samples taken by BESIII~\cite{bes3-white-paper,Li:2021iwf} can be used to measure the decay rates of the intermediate two-body $D^0$ decays, which are important for exploring quark SU(3)-flavor symmetry and its breaking effects,
and thereby benefit the theoretical predictions of $CP$ violation in hadronic $D$ decays~\cite{Saur:2020rgd}.


\section{Acknowledgement}

The BESIII collaboration thanks the staff of BEPCII and the IHEP computing center for their strong support. This work is supported in part by National Key Research and Development Program of China under Contracts Nos. 2020YFA0406400, 2020YFA0406300; National Natural Science Foundation of China (NSFC) under Contracts Nos. 12105076, 11705230, 11625523, 11635010, 11735014, 11822506, 11835012, 11935015, 11935016, 11935018, 11961141012, 12022510, 12025502, 12035009, 12035013, 12061131003, 12192260, 12192260, 12192261, 12192262, 12192263, 12192264, and 12192265; the Chinese Academy of Sciences (CAS) Large-Scale Scientific Facility Program; Joint Large-Scale Scientific Facility Funds of the NSFC and CAS under Contracts Nos. U1732263, U1832207; CAS Key Research Program of Frontier Sciences under Contract No. QYZDJ-SSW-SLH040; 100 Talents Program of CAS; INPAC and Shanghai Key Laboratory for Particle Physics and Cosmology; ERC under Contract No. 758462; European Union's Horizon 2020 research and innovation programme under Marie Sklodowska-Curie grant agreement under Contract No. 894790;; German Research Foundation DFG under Contracts Nos. 443159800, Collaborative Research Center CRC 1044, FOR 2359, GRK 214; Istituto Nazionale di Fisica Nucleare, Italy; Ministry of Development of Turkey under Contract No. DPT2006K-120470; National Science and Technology fund; Olle Engkvist Foundation under Contract No. 200-0605; STFC (United Kingdom); The Knut and Alice Wallenberg Foundation (Sweden) under Contract No. 2016.0157; The Royal Society, UK under Contracts Nos. DH140054, DH160214; The Swedish Research Council; U. S. Department of Energy under Contracts Nos. DE-FG02-05ER41374, DE-SC-0012069.

\end{document}